\documentclass[a4paper,11pt]{article}
\usepackage{aaskaiid,subfigure,orcidlink}
\usepackage[capitalise,noabbrev]{cleveref}
\usepackage{xspace}
\usepackage{ulem}
\usepackage{comment}
\newcommand{\camb}{\texttt{CAMB}\xspace}

\newcommand{\lcdm}{\ensuremath{\Lambda}\rm CDM }

\newcommand{\hi}{\textrm{H\textsc{i}}\xspace}
\newcommand{\twopoint}{\ensuremath{N\times2\mathrm{pt}}\xspace}

\newcommand{\onreview}[1]{{#1}}

\title{Cosmology from \twopoint Analyses of SKAO Wide-Area Surveys}
\ShortTitle{\twopoint Cosmology with SKAO}

\ShortName{Harrison, Sakr et al.} 

\author[1]{Ian Harrison\orcidlink{0000-0002-4437-0770}}
\affiliation[1]{School of Physics and Astronomy, Cardiff University, CF24 3AA, UK}
\emailAdd{harrisoni@cardiff.ac.uk}

\author[2,3,4]{Ziad Sakr\orcidlink{0000-0002-4823-3757}}
\affiliation[2]{Instituto de Física Teórica UAM-CSIC, Campus de Cantoblanco, 28049 Madrid, Spain}
\emailAdd{zsakr@irap.omp.eu}
\affiliation[3]{Institut de Recherche en Astrophysique et Plan\'etologie (IRAP), Universit\'e de Toulouse, CNRS, UPS, CNES, 14 Av. Edouard Belin, 31400 Toulouse, France}
\affiliation[4]{Universit\'e St Joseph; Faculty of Sciences, Beirut, BP-11514, Lebanon}

\author[5,6]{Giulia Piccirilli\orcidlink{0000-0002-3341-1872}}
\affiliation[5]{Dipartimento di Fisica, Università di Roma Tor Vergata, via della Ricerca Scientifica, 1, 00133, Roma, Italy}
\affiliation[6]{INFN - Sezione di Roma 2, Università di Roma Tor Vergata, via della Ricerca Scientifica, 1, 00133 Roma, Italy}

\author[7]{Chandra Shekhar Saraf\orcidlink{0000-0002-5149-4042}}
\affiliation[7]{Korea Astronomy and Space Science Institute, 776 Daedeok-daero, Yuseong-gu, Daejeon 34055, Republic of Korea}

\author[8]{\\Jacobo Asorey\orcidlink{0000-0002-6211-499X}}
\affiliation[8]{Departamento de F\'{\i}sica Te\'orica, Centro de Astropart\'iculas y F\'isica de Altas Energ\'ias (CAPA), Universidad de Zaragoza, 50009 Zaragoza, Spain}

\author[9,10,11]{Benedict Bahr-Kalus\orcidlink{0000-0002-4578-4019}}
\affiliation[9]{INAF –  Osservatorio Astrofisico di Torino, Via Osservatorio 20, 10025 Pino Torinese, Italy}
\affiliation[10]{Dipartimento di Fisica, Universit\`a degli Studi di Torino, Via P.\ Giuria 1, 10125 Torino, Italy}
\affiliation[11]{INFN – Sezione di Torino, Via P.\ Giuria 1, 10125 Torino, Italy}

\author[12,13,14]{José Fonseca\orcidlink{0000-0003-0549-1614}}
\affiliation[12]{Instituto de Astrof\'isica e Ci\^encias do Espa\c{c}o, Universidade do Porto, CAUP, Rua das Estrelas, PT4150-762 Porto, Portugal}
\affiliation[13]{Departamento de F\'isica e Astronomia, Faculdade de Ci\^{e}ncias, Universidade do Porto, Rua do Campo Alegre 687, 4169-007 Porto, Portugal}
\affiliation[14]{Department of Physics and Astronomy, University of the Western
Cape, Robert Sobukwe Road, Cape Town 7535, South Africa}

\author[5,6]{Marina Migliaccio\orcidlink{0000-0003-2040-4654}}

\author[7]{David Parkinson\orcidlink{0000-0002-7464-2351}}

\author[15]{Cora Uhlemann\orcidlink{0000-0001-7831-1579}}
\affiliation[15]{Fakult\"at f\"ur Physik, Universit\"at Bielefeld, Postfach 100131, 33501 Bielefeld, Germany}

\abstract{
SKAO surveys will provide an unprecedented window into the large-scale structure of the universe through HI 21cm galaxy and intensity mapping surveys, and radio continuum surveys. We present forecasts for the cosmological constraining power of ``\twopoint'' analyses -- which combine galaxy clustering, galaxy weak lensing, galaxy-galaxy lensing signals and 21cm Intensity Maps. By assuming cosmology surveys from an SKA-Mid AA4, we show that such an \twopoint analysis will be able to deliver measurements of $\sim1\%$ precision on \lcdm cosmological parameters. We also explore dynamical dark energy in the $w_0, w_a$ model, the sum of neutrino masses $\rm M_\nu$, and the background curvature $\Omega_{\rm k}$.
}

\begin{document}
\maketitle

\section{Introduction}
The three dimensional distribution of matter in the Universe is one of the primary observables in cosmology. How these structures form and grow over cosmic time contains a wealth of information on the dynamics, kinematics and contents of the Universe. In particular, the combination of the cosmological principle (that observers occupy a typical region of a statistically homogenous and isotropic Universe) with the expected Gaussian distribution of initial density perturbations created by inflation, means we most often focus on two-point statistics of astrophysical observables on large areas of the sky. From a three dimensional distribution of matter with a Fourier-space two-point function with a given power spectrum $P(k)$ we observe a spherical projected field with an angular power spectrum $C_\ell$.

Typical examples of these projected fields include galaxy positions (either in full 3D for a spectroscopic survey or tomographic bins for a photometric survey), galaxy shape distortions from gravitational lensing (which traces the gravitational potential) or intensity of spectroscopic line emission (Intensity Mapping). Though all of these projected fields trace the same underlying matter over- and under-densities, each will be affected in different ways by mediating astrophysical and observational processes. Therefore, it has become increasingly profitable to combine multiple tracer spectra in ``\twopoint'' analyses involving both the auto- and cross-power spectra between maps of different tracers, allowing such non-cosmological nuisance effects to be self-calibrated by the combination of different observables.

The classical example of \onreview{this approach} is $3\times2$pt analyses involving clustering in positions of one galaxy sample \onreview{(which here we denote $g$)}, the weak gravitational lensing of a second galaxy sample \onreview{which here we denote $\gamma$} and their cross-correlation, often referred to as `galaxy-galaxy lensing' ($\gamma g$ because it can be thought of as the lensing signal in the second galaxy sample at the positions of the first). In particular, this means that the unknown galaxy bias $b$ for the clustering sample can be informed by the galaxy-galaxy lensing, diminishing the dilution of cosmological information which would otherwise happen due to variation in $b$ being highly degenerate with variation in cosmology. $3\times2$pt analyses of optical and near-Infrared (nIR) photometic galaxy surveys have become one of the most constraining data sets in cosmology \citep[e.g.][]{Heymans:2020gsg,DES:2026fyc}, providing a critical and competitive low-redshift counterpart to the high-redshift Cosmic Microwave Background (CMB). Lensing of the CMB is also increasingly used as a well-understood high-redshift tracer to increase these analyses to $6\times2$pt, both providing further calibration of lensing nuisance parameters and greater sensitivity to physics above $z\sim1$ (e.g., \citealt{DES:2022xxr}). The further addition of maps of the thermal Sunyaev-Zel'dovich (tSZ) effect which traces hot electrons in large scale structure has been shown to signficantly increase the constraining power of the resulting $10\times2$pt \citep{Fang:2023efj}. The power of combined probes to calibrate nuisance parameters and tighten constraints across all probes in shorter-than-radio wavelengths: photometric clustering, spectroscopic clustering, weak lensing, tSZ and CMB lensing has been showcased in the $9\times2$pt with existing data \citep{Reeves:2025axp} and forecast for $12\times2$pt future data \citep{Reeves:2023kjx}.

Within radio astronomy, the SKAO's Mid-frequency telescope \onreview{in its full 197 dish AA4 configuration \citep[][henceforth referred to as Mid-AA4]{seethapuram_sridhar_2025_16951020}} provides the first opportunity to do full-scale \twopoint analyses using radio data. $1\times2$pt analyses of source clustering have long been commonplace, with differences in clustering signal between types of Active Galactic Nuclei (AGN) and star-forming galaxy (SFG) populations being a crucial component of understanding their intrinsic formation and evolution physics. As CMB lensing maps have reached greater signal-to-noise there has also been an increase in cross-correlations between radio galaxy samples and CMB lensing, typically focused on calibrating the bias and redshift distributions of the radio populations rather than cosmology \citep{Alonso:2020jcy,Piccirilli:2022myi,Nakoneczny:2023nlt,Tanidis2025}. Weak lensing surveys remain of interest for SKAO cosmology, and the galaxy-galaxy lensing signal has been detected to high significance in radio data \citep[see][for a detailed consideration]{Harrison02.2026.SKA}.

In addition to galaxy samples, SKA-Mid will be capable of intensity mapping of the HI 21cm emission line \cite[HI IM][]{Wolz01.2026.SKA}, which is again typically analysed with two-point statistics. Cross-correlations \citep{Masui:2012zc,Anderson:2017ert,Cunnington:2022uzo,eBOSS:2021ebm} of HI IM with optical galaxy samples have already led to detections of the signal and useful constraints on the abundance and evolution of cosmological neutral hydrogen, long before the recent detection of the HI IM auto-correlation signal \citep{2025arXiv251119620C,MeerKLASSautodetection}.

In this chapter we discuss the prospects for \twopoint analyses with SKA-Mid in its AA4 configuration. \onreview{In \cref{sec:existing} we provide a brief review of existing cross-correlation measurements involving radio galaxy and intensity mapping large scale structure tracers. In \cref{sec:surveys} we describe the fiducial SKAO wide area surveys which are expected to contribute towards the \twopoint analyses for which we show forecast parameter constraints in \cref{sec:results}. In \cref{sec:cmb_lensing} we also perform a more preliminary forecast, of the detectable signal-to-noise, of the cross-correlation of SKAO observables with CMB lensing data, which is not included in \cref{sec:results}. Finally in \cref{sec:conclusions} we conclude.}

\section{Existing Cross-Correlation Measurements}
\label{sec:existing}
\onreview{To date, SKAO precursor and pathfinder surveys have often measured $1\times2$pt observables, with available signal-to-noise not being high enough to render them suitable for use in \twopoint analyses. They do however provide valuable methodological test beds as well as information on the properties of the radio tracers used in the \twopoint forecasts. In particular, cross-correlations between radio source populations and external tracers have proven constraining on the redshift distribution and linear bias parameters of the radio sources. Modelling and marginalising over uncertainty in these parameters is standard part of \twopoint analyses, meaning prior knowledge from pathfinder and precursor surveys is important to inform our forecasts. Here we provide a brief review of the scope and results of these existing measurements.}

Among recent works, \cite{Saraf:2025wuu} analysed the cross-correlation between radio source positions from the Evolutionary Map of the Universe (EMU; \citealt{EMU_2011}) and photometric optical galaxies from the Dark Energy Survey \citep{DES_2021}, in order to constrain the redshift distribution $n(z)$ of radio galaxies. Along similar lines, \cite{2025arXiv251122732P} investigated the cross-correlation of EMU sources with galaxies from the first data release of the Euclid Deep Field South \cite{Euclid_Q1}. This pathfinder study was able to constrain the bias of the optical galaxies and to characterise the redshift distribution $n(z)$ of radio sources. These works highlight the importance of cross-correlating radio surveys with datasets at different wavelengths to improve the modeling of source properties and, in turn, to obtain tighter constraints on cosmological parameters.

Similarly, the relatively strong foregrounds in \hi IM mean that in current data the signal is typically detected in cross-correlation with high signal to noise maps of optical galaxies in the same redshift range as the IM. Following initial detections \citep{Masui:2012zc,Anderson:2017ert,Cunnington:2022uzo,eBOSS:2021ebm}, this observable has been suggested as a way of learning \onreview{jointly about cosmology and nuisance parameters --} \hi and galaxy bias \citep{Pourtsidou:2016dzn} and the redshift distribution of the optical galaxies \citep{Cunnington:2018zxg} -- as well as a way to remove the instrumental noise from intensity mapping \citep{ShiFeng2020}.

Finally, the lensing effect of the CMB by foreground large-scale structure \onreview{(which we here denote $\kappa$)} has now been detected at a significance of $>40\sigma$ \citep{Planck:2018lbu,2022JCAP-PlanckPR4-lensing,ACT_2023, SPT-3G_2024}. Since this background source emission originates at a very precisely known redshift ($z^{*} \simeq 1100$ in the $\Lambda$CDM model), its lensing kernel is also very well determined. As a result, cross-correlations with foreground galaxy samples can be used both to constrain cosmological parameters and to calibrate nuisance parameters that otherwise enter the kernels.
For SKAO galaxy positions, these nuisance parameters are the degenerate combination of the galaxy bias $b(z)$ and the number density $n(z)$. \cite{Piccirilli:2022myi} measured the cross-power spectrum $C_\ell^{\kappa g}$ between radio galaxies from the \onreview{TIFR GMRT Sky Survey (TGSS, \cite{Intema_2017}}) and \onreview{The NRAO VLA Sky Survey (NVSS, \cite{Condon_1998})} surveys and \textit{Planck} CMB lensing. Similarly, \cite{Nakoneczny:2023nlt} used the same CMB lensing maps in combination with low-frequency LOFAR data from the \onreview{LOFAR Two-metre SkySurvey (LoTSS)} DR2 catalogue (\cite{Shimwell_2022}), constraining both the amplitude and the redshift dependence of the galaxy bias, updating earlier results from \cite{Alonso:2020jcy} based on the LoTSS DR1 sample. The analysis of EMU-PS radio galaxies in \cite{Tanidis2025} showed that the cross-correlation with \textit{Planck} PR4 CMB lensing maps is a robust way to estimate the galaxy bias of radio continuum populations allowing to avoid the systematic induced by the source finders in continuum surveys.  These current analyses achieve signal-to-noise ratios of $\mathcal{O}(20)$ and are already providing useful constraints on radio source properties.

\section{SKAO Cosmology Surveys}
\label{sec:surveys}
The combination of survey speed and sensitivity from SKA-Mid AA4 will make it the first radio experiment capable of \onreview{achieving signal-to-noise high enough across multiple observables to perform full-scale, cosmologically constraining \twopoint analyses at radio frequencies. In designing these surveys there will be an} optimal choice of trade-off between depth and area for a given amount of observing time (typically \onreview{assumed to be} 10,000 hours as consistent with expectations for large SKAO programmes). \onreview{This optimisation is discussed in \cite{SKA:2018ckk} and extensively in the references therein. Here we recount the key features of each notional survey}.

\subsection{Continuum Galaxy Clustering and Weak Lensing}
\label{sec:continuum}
\onreview{Radio continuum galaxy surveys detect the flux of distant galaxies integrated over a wide frequency band; their use for cosmology is covered in the volume by \cite{Asorey01.2026.SKA}. }For a continuum galaxy survey we consider an AA4 survey which provides both galaxy positions and galaxy shapes from the same overall sample (as opposed to separate `lens' and `source' samples of different galaxies)\onreview{, consistent with the \textit{Medium-Deep Band 2 Survey} of \cite{SKA:2018ckk}. We note that the sensitivity estimates used at that time \citep[from][]{braun2019anticipatedperformancesquarekilometre} remain in line with those from the current sensitivity calculator for the AA4 configuration\footnote{\url{https://sensitivity-calculator.skao.int/}}}. This survey has sky area $A_{\rm sky} = 5,000 \, \mathrm{deg}^2$ \onreview{(expected to coincide with the DES survey footprint)}, number density of galaxies with shape measurements $n_{\rm gal} = 2.7\,\mathrm{arcmin}^{-2}$, median redshift $z_{\rm m} = 1.1$ and shape noise dispersion $\sigma_\epsilon = 0.3$. 

For galaxy clustering the galaxy bias is assumed to be \onreview{linear and }constant within each redshift bin $i$, and its values $b(z_i)$ (a total of ten parameters) are marginalised over as nuisance parameters. Similarly, for the weak lensing case we marginalise over a three parameter \onreview{(redshift evolving Nonlinear Linear Alignment)} model for galaxy intrinsic alignments (\onreview{NLA-}IA) which can contaminate the cosmological signal \citep[see][and references therein for a discussion of IA in radio weak lensing]{Harrison02.2026.SKA}. For both the galaxy clustering signal and the weak lensing signal, we use a multipole range $10 < \ell < 1000$. \onreview{The large scale cut at $l=10$ is motivated by the control of large scale systematics including galactic contamination, as discussed in \cite{Asorey01.2026.SKA}.} \onreview{Within limit of the small-scale cut, our choice of only including linear galaxy bias is a mildly optimistic one, with the $\ell$-space analysis including some non-linear scales at higher redshifts. However, the choice of how to model these non-linear scales is a complicated one outside of the scope of these indicative forecasts \citep[see e.g.][for an extensive discussion]{Euclid:2026glh}.}

We consider ten equally populated tomographic bins with the redshift distribution of sources $n(z)$ \onreview{following} the well-known \cite{Smail:1994sx} functional form (which we evaluate over a redshift range $0 < z < 5$):
\begin{equation}
    n(z) = n_{\rm gal}\,\left( \frac{z}{z_{\rm m}/\sqrt2} \right)^2\,\exp\left[-\left( \frac{z}{z_{\rm m}/\sqrt2} \right)^{3/2}\right]\;,
	\label{eq:redshift distribution}
\end{equation}
and with photometric errors $\sigma_z = 0.05(1 + z)$, with the redshift estimation expected to come from cross-matching sources with optical and nIR surveys. \onreview{The provision of redshift information for sources in deep radio continuum observations is an ongoing area of investigation, and is not explored in detail here. For the bulk of sources at low redshifts $z \lesssim 2$ it is expected they will have matched counterparts in optical and nIR surveys, principally LSST \citep[recent studies using multi-wavelength data to provide redshifts for radio sources in LoTSS have found cross-matches in the 60-90\% range][]{Kondapally:2020nym,2023A&A...678A.151H}. Further optimisation of the binning scheme may improve the forecasts here, but we consider it beyond the scope of this paper.}

\onreview{Further calibration of the shape of the radio redshift distribution, including in in the higher redshift regime, can also be gained through the dependence of a parametric $n(z)$ model on the $1\times2$pt measurement between optical galaxy and CMB lensing data, as performed in the references discussed in \cref{sec:existing} above. However, the signal-to-noise will be high enough with future surveys to measure a non-parametric reconstruction of the $n(z)$, maximising the amount of information used for the redshift distribution. To study the feasibility of such model independent reconstructions, we prepared forecasts for the cross-correlation of SKA-Mid AA4 and Euclid-like survey galaxy positions using \texttt{GLASS} \citep{2023OJAp....6E..11T}.}

\begin{figure}
    \centering
    \includegraphics[width=0.5\linewidth]{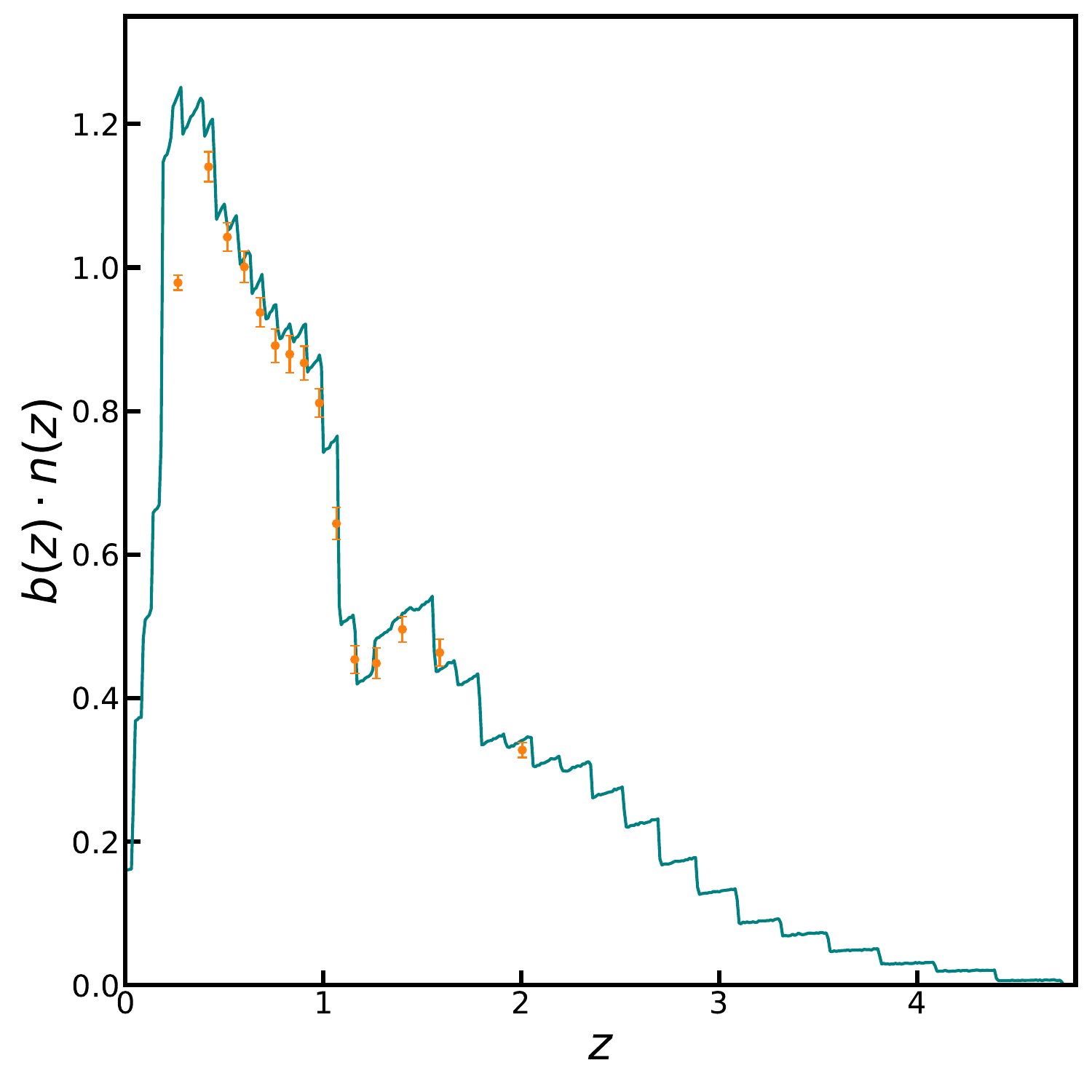}
    \caption{Reconstruction of SKA-Mid window function from cross-correlation with Euclid-like galaxy survey. The green solid line is the fiducial SKA-Mid window function and the orange points denote the recovered distribution.}
    \label{fig:ska_window}
\end{figure}

\onreview{For SKA-Mid we use the \textit{Medium-Deep Band 2} survey specifications as described here, while for \textit{Euclid}-like survey we assumed a $\sim 16,000$ deg$^{2}$ sky area coverage with source density  $n_{\text{gal}} = 30$ arcmin$^{-2}$ and a redshift distribution modelled using Eq. \ref{eq:redshift distribution} with $z_{m} = 0.9$. The \textit{Euclid} galaxies were divided into 15 equi-density bins. In Figure \ref{fig:ska_window}, we show the SKA-Mid galaxy window function (the degenerate combination of $n(z)$ and $b(z)$ the cross-correlation is sensitive to) inferred from cross-correlation spectra alone. We note that the window function is recovered within $1\sigma$ up-to $z\sim 2$, the redshift range of \textit{Euclid} galaxies.}

\subsection{HI Intensity Mapping}
\label{sec:hiim}
\onreview{Neutral Hydrogen (\hi) surveys detect the line emission from the 21cm spin-flip transition, with \hi intensity mapping (IM) surveys \cite[see][in this volume]{Wolz01.2026.SKA} detecting the low spatial resolution emission from many galaxies in a single beam.}For a neutral Hydrogen intensity mapping survey with Mid AA4 we consider a wide area single-dish mode survey \onreview{consistent with the \textit{Wide Band 1 Survey} of \cite{SKA:2018ckk}, } with $A_{\rm sky} = 20,000 \, \mathrm{deg}^2$ and a system noise level as given in Table 7. of \cite{Jolicoeur:2020eup}. We include information from a HI power spectrum on scales $0.001 < k \, \mathrm{Mpc}^{-1} < 0.15$, with the HI density and bias given by:
\begin{align}
\label{eq:hi_bias}
    b_{\rm HI}(z) &= c_{\mathrm{HI},1} (1 + z) + 0.5 \\
\label{eq:hi_density}
    \Omega_{\rm HI}(z) &= 4.0 (1 + z)^{c_{\mathrm{HI},2}} \times 10^{-4},
\end{align}
with $c_{\mathrm{HI},1} = 0.3$ and $c_{\mathrm{HI},2} = 0.6$ in line with accepted models for existing data \citep{Switzer:2013ewa,Crighton:2015pza}. Though SKAO will be able to observe in significantly narrower frequency channels, cosmological information can be usefully extracted from much coarser binning. Here we use ten redshift bins for \hi IM, equally spaced in the range $\left[ 0.5, 2.6\right]$.

\subsection{HI Galaxy Clustering}
\label{sec:higal}
\onreview{\hi 21cm line emission can also be detected in spatially resolved galaxies, whose three dimensional positions and velocities may then be used for cosmology \citep[see][in this volume for a detailed discussion]{Nasirudin01.2026.SKA,Mayor01.2026.SKA}.} For a spectroscopic galaxy clustering survey we consider a \hi spectral line survey with $A_{\rm sky} = 5,000 \, \mathrm{deg}^2$\onreview{, again consistent with the \textit{Medium-Deep Band 2 Survey} of \cite{SKA:2018ckk}}. Following \cite{Ronconi01.2026.SKA} we assume a source number density and bias model for the galaxies:
\begin{align}
    \frac{dN}{dz} &= 10^{c_1} z^{c_2} \exp \left( -c_3 z \right) \\
    \label{eq:hi_gal_bias}
    b(z) &= c_4 \exp \left( c_5 z \right)
\end{align}
with $\lbrace c_1, c_2, c_3, c_4, c_5 \rbrace = \lbrace 5.63, 1.41, 15.49, 0.605, 1.086 \rbrace$. We assume a spectroscopic error for the galaxies redshift uncertainty of $\sigma_z = 0.002$ and bin the spectroscopic galaxies into five equally populated redshift bins with redshift $\left[ 0.0, 0.5\right]$. As for IM we include information from the power spectrum on scales $0.001 < k \, \mathrm{Mpc}^{-1} < 0.15$, also a conservative choice suitable to our adopted semi-analytic nonlinear redshift space distortion. This model consists of a Kaiser correction as well as considering free terms for each bin, \onreview{which} account for the nonlinear Fingers-of-God and BAO damping\onreview{, for which we include parameters ($\sigma_{p, i}$ and $\sigma_{v,i}$ respectively) for each bin which are marginalised over, as listed in \cref{tab:params}}. \onreview{Again, here we note our treatment of nonlinear clustering may be regarded as somewhat optimistic; in reality marginalisation over additional non-linear bias parameters may increase the inferred uncertainties.}

\subsection{External Data}
\label{sec:external_data}
In order to constrain parameters to which the low redshift large scale structure observables from SKA-Mid are not very sensitive, we also include primary CMB and CMB lensing spectra data. Both of these we take to be observed at the \textit{Planck} 2018 level of precision, specifically \cite{Planck:2018nkj,Planck:2019nip,Planck:2018vyg} for TT, TE and EE spectra, and \cite{Planck:2018lbu} for lensing. We also vary the optical depth to reionisation $\tau$ that we marginalize over later when we combine with SKAO data. Although better small scale CMB data will be available from the Simons Observatory Large Aperture Telescope \citep[SO-LAT][]{SimonsObservatory:2018koc,SimonsObservatory:2025wwn} contemporaneous with AA4 surveys, the differential in precision on the cosmological parameters of interest will be relatively small when compared to that between SKAO with and without primary CMB.

\section{Cosmology from SKAO Observations}
\label{sec:results}
\begin{table}
	\centering	
	\caption{	
         Unlisted other cosmological parameters and model choices are fixed to their default values in \camb v1.3.5 \citep{Lewis:1999bs}}
	\vspace{-0.2cm}
	\begin{tabular}{ccl}
		\hline
            \hline
		Parameter & Fiducial Value & Description \\\hline
		\multicolumn{3}{l}{\textbf{Cosmology}}  \\
		$\Omega_{\rm m}$ & 0.320 & Total matter density \\ 
        $\Omega_{\rm b}$ & 0.050 & Baryonic matter density \\
        $h$ & 0.67 & Hubble parameter \\
        $n_s$ & 0.960 & Scalar spectral index \\
        $\sigma_8$ & 0.816 & Amplitude of fluctuations \\
        $\tau$ & 0.058 & Optical depth to reionisation \\
        $w_0$ & $-1.0$ & Dark Energy equation of state \\
        $w_a$ & 0.0 & Dark Energy equation of state rate of change \\
        $\Omega_k$ & 0.0 & Curvature density \\
        $\sum m_{\nu} \, [\mathrm{eV}]$ & 0.06 & Sum of standard neutrino masses \\
		\hline

  
  		\multicolumn{3}{l}{\textbf{Continuum Galaxy Nuisance} }	 \\
		$A_{\rm IA}$ & 1.72 & Intrinsic Alignment (IA) NLA amplitude \\
            $\beta_{\rm IA}$ & 2.17 & IA NLA redshift evolution\\
            $\eta_{\rm IA}$ & $-0.41$ & IA NLA luminosity function dependence \\
            $b_i , i \in [1,10]$ & $[1.10, 1.22, 1.127, 1.31, 1.36,$ & Galaxy linear bias\\
             & $ 1.40, 1.44, 1.49, 1.56, 1.74]$ & \\
            \hline
        \multicolumn{3}{l}{\textbf{\hi Galaxy Nuisance} }	 \\
            $b_{\mathrm{HI},i} , i \in [1,5]$ & $[0.65, 0.72, 0.79, 0.88, 0.97]$ & \hi Galaxy linear bias\\
            $P_{S,i} , i \in [1,5]$ & $[0.0, 0.0, 0.0, 0.0, 0.0, 0.0]$ & \hi Galaxy shot noise\\
            $\sigma_{p,i} , i \in [1,5]$ & $[1.0, 1.0, 1.0, 1.0, 1.0, 1.0]$ & FoG pairwise velocity dispersion\\
            $\sigma_{v,i} , i \in [1,5]$ & $[1.0, 1.0, 1.0, 1.0, 1.0, 1.0]$ & \hi Galaxy BAO damping\\
            \hline
        \hline
        \multicolumn{3}{l}{\textbf{\hi Intensity Mapping Nuisance} }	 \\
            $c_{\mathrm{HI},1}$ & $0.3$ & \hi bias amplitude \cref{eq:hi_bias} \\
            $c_{\mathrm{HI},2}$ & $0.6$ & \hi density redshift evolution \cref{eq:hi_density} \\
            \hline

		

		\hline
	\end{tabular}
	\label{tab:params}

\end{table}
\onreview{We now construct forecasts using the Fisher Matrix formulation in angular power spectrum space, assuming Gaussian likelihoods. For a full description of the formalism used see Section 3 of \cite{Casas:2022vik}.}Here we consider several combinations of \twopoint data from SKA-Mid:
\begin{itemize}
    \item `Cont.' consisting of $3\times2$pt from continuum galaxy clustering (GC, $C_\ell^{gg}$), galaxy weak lensing (WL, $C_\ell^{\gamma\gamma}$) and their cross-correlation (GGL, $C_\ell^{g\gamma}$), as described in \cref{sec:continuum}. This includes three nuisance parameters for galaxy intrinsic alignments and ten nuisance parameters for linear galaxy bias.
    \item `HI' consisting of spectroscopic observations of the power spectrum as probed by HI Intensity Maps (IM) as described in \cref{sec:hiim} and HI galaxy clustering as described in \cref{sec:higal} (Gal). \onreview{This includes the additional twenty-two parameters for \hi galaxy and intensity mapping observables.}
    \item `All' consisting of the combination of the Continuum and HI sets into a $5\times2$pt combination. \onreview{We do not include the other cross-correlations which could be included up to $10\times2$pt. In order to remove diffuse foreground effects, \hi IM often have regions of $k$-space excised which are precisely those which would carry the cross-correlation signal with photometric galaxy redshift surveys \citep{Witzemann:2018cdx}. Though this information could be recovered in the case of sufficiently accurate foreground information, or with bispectrum modelling \citep{Guandalin:2021sxw}, these considerations are beyond the scope of this paper and we chose not to include them. Similarly for \hi intensity maps and spectroscopic galaxies we may expect a complicated covariance which we choose to avoid modelling.}
\end{itemize}
\onreview{The full list of cosmological and nuisance parameters used in shown in \cref{tab:params}.} For the `all' combination we also cumulatively add the \textit{Planck} 2018 Legacy primary CMB data (`P18') and its lensing (`lens'). \onreview{We do not include the cross-correlation terms with CMB data here, but discuss the prospect of their use in \cref{sec:cmb_lensing}.}

\onreview{As an external experiment for comparison of our results in \cref{tab:percent_rel_errors_lcdm,tab:percent_rel_errors_w0wa,tab:percent_rel_errors_mnu} we use the Fisher forecasts for the \textit{Euclid} experiment from \cite{Euclid:2019clj} which include spectroscopic galaxy clustering (GC$_{\rm s}$), photometric galaxy clustering (GC$_{\rm Ph}$), galaxy weak lensing (WL) and the the galaxy-galaxy lensing cross-correlation (XC$^{(\mathrm{GC}_{\rm Ph}, \mathrm{WL)}})$}
\subsection{Constraints in \lcdm}
\begin{figure}[h]
    \centering
	\includegraphics[width=\columnwidth]{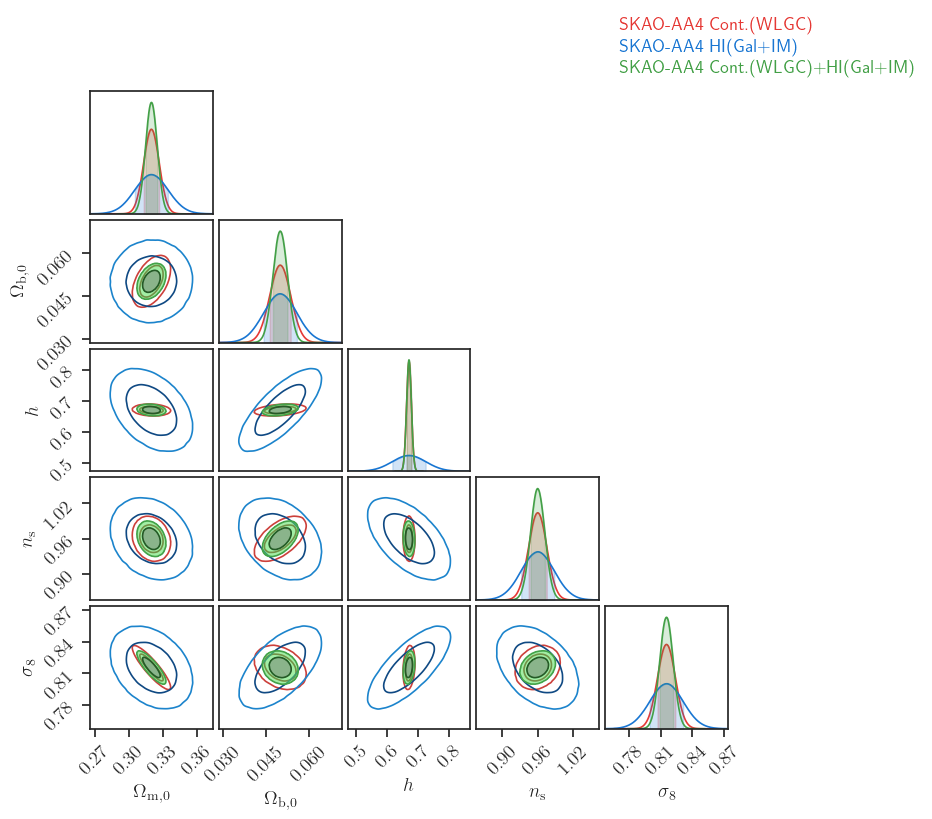}
    \caption{The constraining power of \twopoint combinations of SKA-Mid AA4 data. Red shows the $3\times2$pt combination of continuum galaxy clustering, weak lensing and galaxy-galaxy lensing. Blue shows the additive combination of $1\times2$pt from HI galaxies and $1\times2$pt from HI Intensity Mapping (IM). Green shows the combination of all of these in a $5\times2$ combination. Note in particular certain combinations where the different degeneracy directions (principally the $\Omega_{\rm b,0}$ column) enable significantly tighter constraints in combination.
    }
    \label{fig:skao_nx2_lcdm}
\end{figure}

We first show results for the parameters of the \lcdm model,  \Cref{tab:percent_rel_errors_lcdm} shows the relative percent errors on the parameters from \twopoint combinations, with the full Fisher information including probe covariances shown in \cref{fig:skao_nx2_lcdm}. As can be seen, the different degeneracy directions of the continuum and HI observables in a number of panels (different parameter combinations) acts to reduce the error regions significantly. For the combination of Cont. and HI all parameters other than the baryon density $\Omega_{\rm b,0}$ are constrained to the sub-percent level, which is currently the gold-standard from CMB experiments at high redshift. \onreview{When compared to the \textit{Euclid} forecasts, we can see the power of SKAO is less by factors $\sim2 $-$10$, depending on the parameter. This matches the existing expectation that SKAO in its AA4 configuration will sit between Stage III and Stage IV-like cosmology surveys, as defined by the Dark Energy Task Force document \citep{Albrecht:2006um}. As discuss in \cite{SKA:2018ckk} the proposed SKAO Phase 2 telescope, with greater sensitivity and angular resolution, will have capabilities more in line with Stage IV surveys such as \textit{Euclid}.}

\begin{table}[ht]
\small
\begin{center} 
\begin{tabular}{ccccccc} 
\hline 
Case & $\Omega_{{\rm m},0}$ & $\Omega_{{\rm b},0}$ & $h$ & $n_{\rm s}$ & $\sigma_8$ \\
\hline
\multicolumn{1}{c}{\textbf{\boldmath{$\Lambda$}CDM}}\\
\hline
SKAO: Cont.~(WLGC) & 2.10~\% & 7.22~\% & 1.15~\% & 1.59~\% & 1.02~\%\\
SKAO: HI~(Gal$+$IM) & 4.52~\% & 11.52~\% & 7.95~\% & 2.89~\% & 1.91~\%\\
SKAO: All (Cont.$+$HI) & 1.59~\% & 5.01~\% & 1.13~\% & 1.25~\% & 0.77~\% \\
\onreview{\textit{Euclid} (GC$_{\rm S}$+WL+GC$_{\rm Ph}$+XC) Pess.} & 0.7~\% & 2.6~\% & 0.37~\% & 0.53~\% & 0.33~\% \\
\onreview{\textit{Euclid} (GC$_{\rm S}$+WL+GC$_{\rm Ph}$+XC) Ope.} & 0.25~\% & 1.1~\% & 0.11~\% & 0.15~\% & 0.11~\% \\
CMB (P18) + SKAO (all) & 0.55~\% & 0.35~\% & 0.18~\% & 0.17~\% & 0.20~\% &\\
CMB + lens (P18) + SKAO (all) & 0.30~\% & 0.21~\% & 0.10~\% & 0.11~\% & 0.09~\% \\
\hline
\end{tabular}
\end{center}
\caption{Showing 1~$\sigma$ percentage relative errors on the \lcdm parameters for different \twopoint cases, where `all' refers to the combination of WLG and G\&IM. \onreview{The \textit{Euclid} rows refer to the optimistic and pessimistic scenarios from Table 9 of \cite{Euclid:2019clj}.}}
\label{tab:percent_rel_errors_lcdm}
\end{table}

\subsection{Constraints in $w_0 w_a$\rm CDM}
\begin{figure}[h]
    \centering
	\includegraphics[width=\columnwidth]{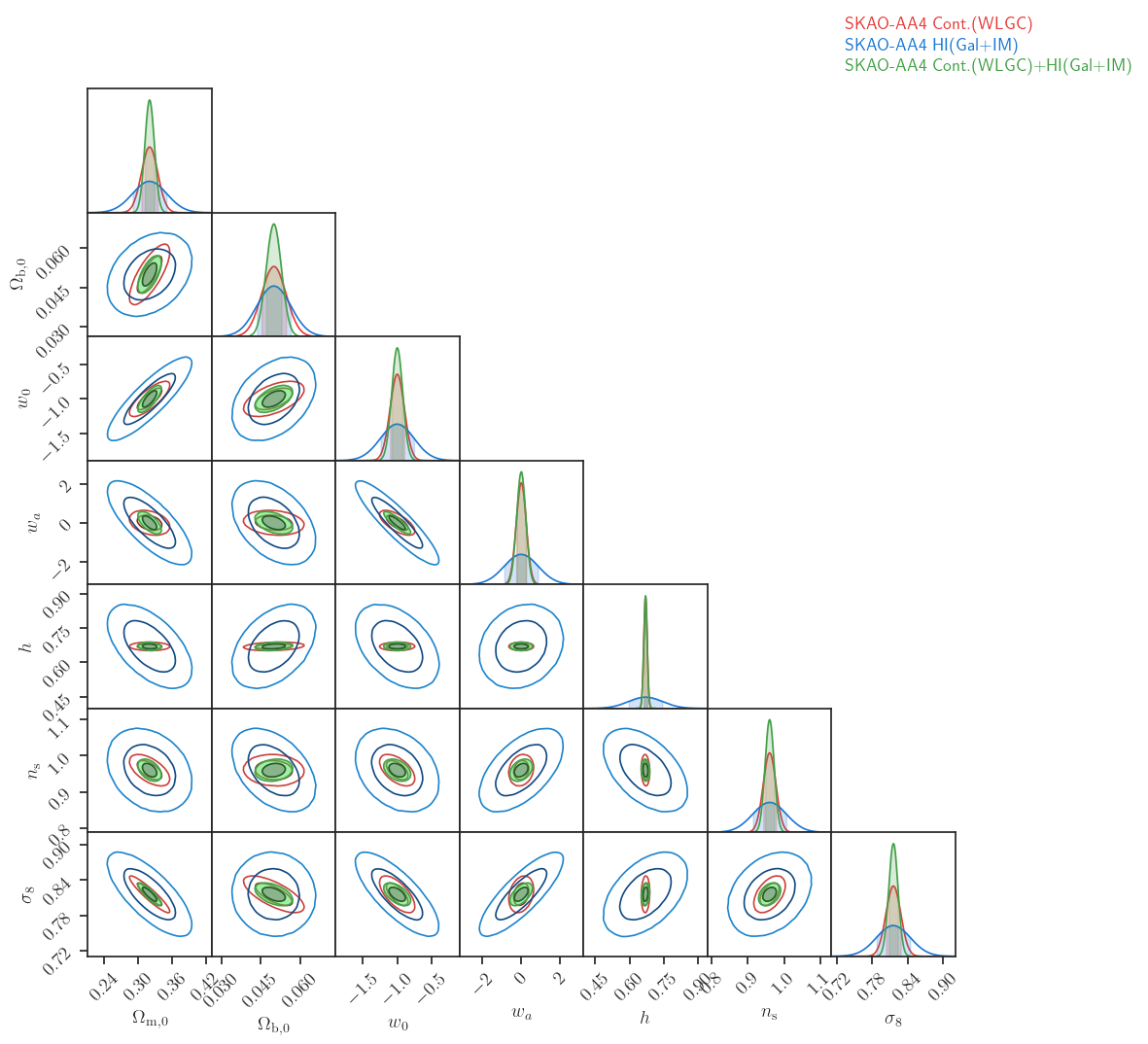}
    \caption{As \cref{fig:skao_nx2_lcdm} but for the case where $w_0, w_a$ dark energy parameters are also included.}
    \label{fig:skao_nx2_w0wa}
\end{figure}

In order to consider a dynamical dark energy model, we extended our cosmological parameter set to include $w_0$ and $w_a$ from the Chevallier-Polarski-Linder \citep[CPL; ][]{Chevallier:CPL,Linder:CPL} parameterisation of dark energy. \Cref{tab:percent_rel_errors_w0wa} and \cref{fig:skao_nx2_w0wa} show the results when these are varied alongside the \lcdm parameters. As can be seen the additional degeneracies increase the fractional errors on all parameters, especially the total matter content $\Omega_{\rm m,0}$. Of particular importance is the combination of the different redshift ranges of Continuum and HI probes, which are vital in reducing the otherwise large uncertainties on the $w_0, w_a$ parameters in particular. \onreview{As for the pure \lcdm case, the parameter uncertainties here are factors $\sim$2-10 greater than those for \textit{Euclid}.}

\begin{table}[ht]
\small
\begin{center} 
\begin{tabular}{cccccccc} 
\hline 
Case & $\Omega_{{\rm m},0}$ & $\Omega_{{\rm b},0}$ & $h$ & $n_{\rm s}$ & $\sigma_8$ & $w_0$ & $w_a$ \\
\hline
\multicolumn{1}{c}{\textbf{\boldmath{$w_0w_a$CDM}}}\\
\hline
SKAO: Cont.~(WLGC) & 4.45~\% & 9.22~\% & 1.11~\% & 1.82~\% & 1.53~\% & 10.29~\% & 25.45~\% \\
SKAO: HI~(Gal$+$IM) & 9.34~\% & 12.69~\% & 10.93~\% & 4.81~\% & 3.49~\% & 24.33~\% & 85.73~\%  \\
SKAO: All (Cont.$+$HI) & 2.60~\% & 5.73~\% & 1.09~\% & 1.28~\% & 0.95~\% & 7.87~\% & 22.99~\%  \\
\onreview{\textit{Euclid} (GC$_{\rm S}$+WL+GC$_{\rm Ph}$+XC) Pess.} & 1.2~\% & 3.6~\% & 0.55~\% & 0.57~\% & 0.51~\% & 4.0~\% & 17~\% \\
\onreview{\textit{Euclid} (GC$_{\rm S}$+WL+GC$_{\rm Ph}$+XC) Opt.} & 0.57~\% & 1.5~\% & 0.15~\% & 0.19~\% & 0.21~\% & 2.5~\% & 9.2~\% \\
CMB (P18) + SKAO (all) & 1.73~\% & 1.71~\% & 0.85~\% & 0.17~\% & 0.64~\% & 5.94~\% & 17.92~\% \\
CMB + lens (P18) + SKAO (all) & 1.50~\% & 1.49~\% & 0.73~\% & 0.11~\% & 0.54~\% & 3.95~\% & 11.66~\% \\
\hline
\end{tabular}
\end{center}
\caption{As in \cref{tab:percent_rel_errors_lcdm}, but for the case where $w_0, w_a$ dark energy parameters are also included. \onreview{The \textit{Euclid} rows refer to the optimistic and pessimistic scenarios from Table 11 of \cite{Euclid:2019clj}.}}
\label{tab:percent_rel_errors_w0wa}
\end{table}

\subsection{Constraints in $\lcdm+M_\nu$}
\begin{figure}[h]
    \centering
	\includegraphics[width=\columnwidth]{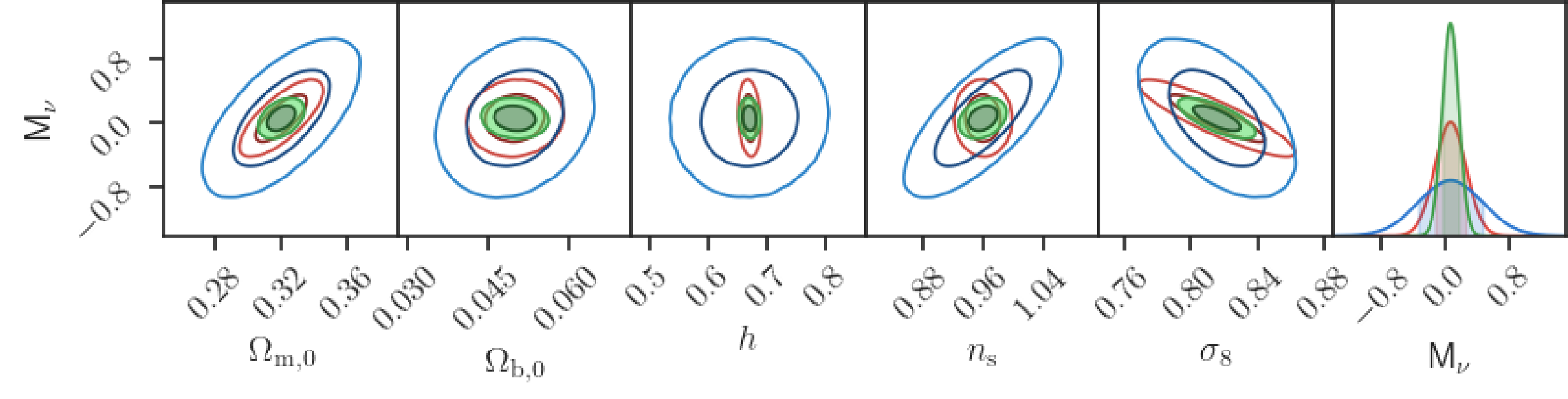}
    \caption{The constraining power of \twopoint combinations of SKA-Mid AA4 data, as in \cref{fig:skao_nx2_lcdm} but here including the constraint on the sum of neutrino mass $\rm M_\nu$.}
    \label{fig:skao_nx2_mnu}
\end{figure}

Finally, we consider the case of the \lcdm parameters from above, with the addition of the sum of neutrino masses $\rm M_\nu$, which may be expected to affect the SKAO observables through both its effect on background expansion, and the formation of structure. \cref{tab:percent_rel_errors_mnu} and \cref{fig:skao_nx2_mnu} show the constraints on $\rm M_\nu$, along with the \lcdm parameters. Here again, the combination of SKAO observables in the full $5\times2$pt with continuum and HI surveys improves by a factor 2 on what is possible with one or the other observables alone.

\begin{table}[ht]
\small
\begin{center} 
\begin{tabular}{ccccccc} 
\hline 
Case & $\Omega_{{\rm m},0}$ & $\Omega_{{\rm b},0}$ & $h$ & $n_{\rm s}$ & $\sigma_8$ &  M$_\nu$ \\
\hline
\multicolumn{1}{c}{\textbf{\boldmath{$\nu \Lambda$}CDM}}\\
\hline
SKAO: Cont.~(WLGC) & 3.14~\% & 7.22~\% & 1.16~\% & 1.59~\% & 2.30~\% & 317.27~\% \\
SKAO: HI~(Gal$+$IM) & 5.94~\% & 11.80~\% & 7.97~\% & 4.31~\% & 2.30~\% & 653.18~\% \\
SKAO: All (Cont.$+$HI) & 1.74~\% & 5.02~\% & 1.13~\% & 1.30~\% & 1.13~\% & 169.16~\% \\
CMB (P18) + SKAO (all) & 1.52~\% & 1.10~\% & 0.57~\% & 0.17~\% & 0.92~\% & 55.87~\% \\
CMB + lens (P18) + SKAO (all) & 1.14~\% & 0.79~\% & 0.41~\% & 0.12~\% & 0.43~\% & 32.14~\% \\
\hline
\end{tabular}
\end{center}
\caption{As in \cref{tab:percent_rel_errors_lcdm}, but for the case where the $\rm M_\nu$ neutrino mass parameter is also included.}
\label{tab:percent_rel_errors_mnu}
\end{table}

\subsection{Constraints in $\lcdm+\Omega_{\rm k}$}
\begin{figure}[h]
    \centering
	\includegraphics[width=\columnwidth]{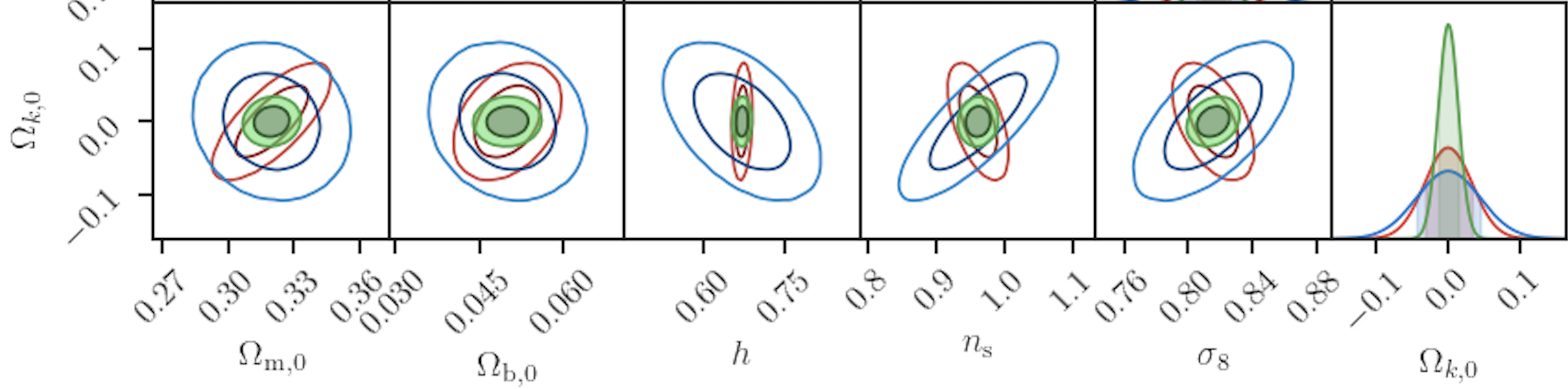}\\
    \includegraphics[width=\columnwidth]{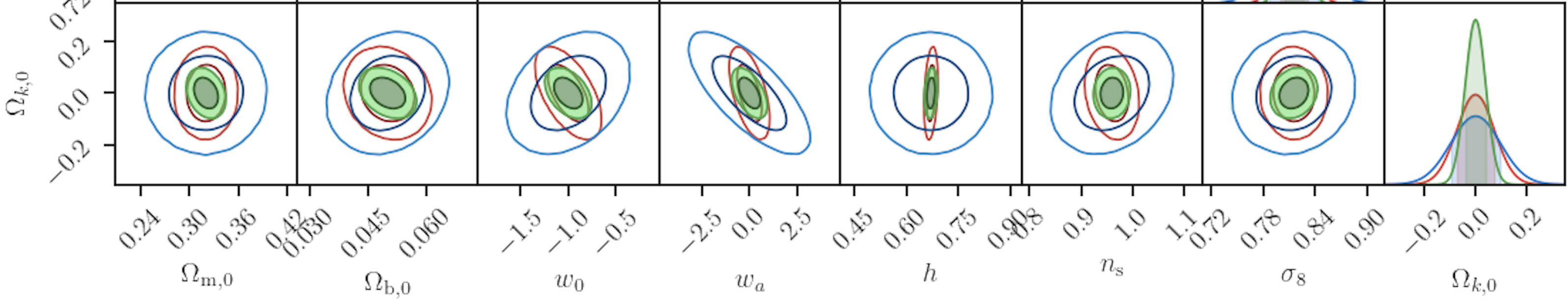}
    \caption{The constraining power of \twopoint combinations of SKA-Mid AA4 data, as in \cref{fig:skao_nx2_lcdm} but here including the constraint on the background curvature $\Omega_{\rm k}$ \onreview{on top of the  \lcdm (\textit{top}) or $w_0,w_a$ (\textit{bottom}) parameters}.}
    \label{fig:skao_nx2_omk}
\end{figure}

As an additional parameter combination we also consider non-flat \lcdm cosmologies by including variation of the $\Omega_{\rm k}$ curvature parameter (we could equivalently vary the dark energy density parameter $\Omega_{\mathrm{DE}, 0}$). The background curvature may have been measured to be non-zero, but typically remains consistent with zero if other possible extensions are also included \citep[see e.g.][section 4.1 for further discussion]{DESI:2024mwx}. Here, as can be seen in \cref{fig:skao_nx2_omk} again the combination of SKA-Mid continuum and HI probes provides useful degeneracy breaking improving the constraints on $\Omega_{\rm k}$ by a factor $\sim 3$.

\section{Cosmology from SKAO and CMB Observables}
\label{sec:cmb_lensing}
\onreview{As discussed in \cref{sec:existing}, the lensing map of the CMB provides a well-measured tracer of large scale structure at a very high redshift. Because CMB lensing also has essentially no nuisance parameters, being from a very precisely known redshift and with high robustness to contamination by non-lensing signals, when combined with other tracers it can be extremely powerful in calibrating their nuisance parameters and breaking degeneracies on cosmological parameters.}

\onreview{However, because it traces the same large scale structure as the low redshift SKAO probes it may be expected to have significant correlations when included together in \twopoint analyses, which is usually characterised with simulations \citep[see e.g.][]{ACT:2023oei}. Accurately characterising this is beyond the scope of this work \citep[see][for an example calculation]{Kou:2025hvg}, and to be conservative we decide not to include the CMB lensing cross-correlations in our forecasts and instead consider only the internal \twopoint of SKAO probes. However, we do carry out an initial analysis of the signal-to-noise at which $1\times 2$pt combinations of CMB lensing with SKAO surveys could be measured.}

\onreview{As an initial forecast we predict the} signal-to-noise ratio ($S/N$, hereafter) for the cross-correlation between the positions of continuum galaxies and CMB probes, such as the temperature anisotropies and the lensing convergence. For the continuum radio sources, we use the redshift distribution and bias model as described in \cref{sec:continuum}.
As mentioned in \cref{sec:external_data}, we consider two CMB datasets: the full-sky existing \textit{Planck} data and the future data which will be available from the Simons Observatory (SO). We use the publicly available noise expectation for the original SO-baseline experiment as described in \cite{SimonsObservatory:2018koc} but note that these will be improved upon by the approved `Enhanced SO' experiment described in \cite{SimonsObservatory:2025wwn}.

We compute the fiducial angular power spectra using the \textsc{CAMB} code\footnote{\url{https://camb.readthedocs.io/en/latest/}}, assuming a $\Lambda$CDM cosmology with parameters from \cite{Planck:2018lbu} and including general relativistic effects \citep{Challinor_2011,Bonvin&Durrer}. The SNR can then be assessed as
\begin{equation}
\label{eq:s2n}
\Bigg(\frac{{\rm S}}{{\rm N}}\Bigg)^2_\ell=\sum_{\ell'}
C^{\rm XY}_\ell \left[{\rm Cov}^{-1}\right]_{\ell\ell'}
 C^{\rm XY}_{\ell'},
\end{equation}
where ${\rm Cov}$ is the Gaussian covariance matrix. In our analysis, ${X,Y} \in \{\kappa, T, g \}$, where $\kappa$ denotes the CMB lensing convergence, $T$ the CMB temperature anisotropies, and $g$ the continuum galaxy overdensity.

The cumulative SNR over a multipole range $[\ell_{\rm min}, \ell_{\rm max}]$ is given by
\begin{equation}
\label{eq:s2n_tot}
\frac{{\rm S}}{{\rm N}} = 
\sqrt{\sum_{\ell=\ell_{\rm min}}^{\ell_{\rm max}}
\Bigg(\frac{{\rm S}}{{\rm N}}\Bigg)^2_\ell}.
\end{equation}

As in \cref{sec:continuum}, we set $\ell_{\rm max}=1000$ and extend the range to $\ell_{\rm min}=2$, since we expect cross-correlations to be less sensitive to spurious signals that may artificially boost the auto angular power spectra on large scales.
We find $(S/N)_{Tg} = [2.9;2.6]$ and $(S/N)_{\kappa g} = [40.8;68.9]$ for \textit{Planck} and SO, respectively.

\onreview{As in discussed in \cref{sec:continuum} when performing the forecast for \cref{fig:ska_window}, this cross-correlation of SKAO galaxy positions with CMB lensing depends on a degenerate combination of redshift distribution and bias of the radio sources. A potential way to break this degeneracy is to use the weak lensing signal of the galaxies rather than their positions, as advocated in \cite{Kalaja:2024tsk}. This would also allow constraints on cosmological parameters coming from a set of angular scales and redshift ranges that are intermediate between the two individual lensing probes, potentially providing useful information on the physics of structure formation in those regimes.}

\section{Conclusions}
\label{sec:conclusions}
In addition to the individual cosmological constraints which can be gained from analyses of individual SKAO cosmology surveys, we have shown the benefits of combining the Continuum observations of galaxy clustering and weak lensing  with the spectroscopic HI observations of galaxies and intensity maps. The different dependencies of these observables on redshift and physical scales mean they are capable of breaking degeneracies between cosmological parameters. This is of particular importance for constraining extended parameter models such as for dark energy and the total neutrino mass.

Such \twopoint analyses are also capable of calibrating nuisance parameters such as unknown galaxy redshift distributions, linear galaxy bias, weak lensing shear bias and intrinsic alignments,  leading to more robust cosmological results. This multiple cross-correlation approach has become the standard in cosmological analyses of current and next-generation experiments, and SKAO will be capable of competitive measurements under the assumption that large-scale mapping of continuum and HI observables is performed.

We emphasise that any cosmological relevance relies on the promise of AA4. Although some loss of sensitivity in the \onreview{intermediate AA$^*$ 144-dish} configuration can be recovered with greater observing time for galaxy clustering and \hi observables, the detection of weak lensing of continuum sources is simply not possible without the long baselines only included in AA4. Without the lensing observables, all of the cosmological parameters here become strongly dependent on the unknown galaxy linear bias function $b(z)$. Though this may be calibrated to some level with cross-correlations with lensing external to SKAO (e.g. CMB or optical surveys), the signal-to-noise available will be far lower, limiting the level of calibration available.

However, in the case where wide-area surveys of both radio continuum and HI are available from SKA-Mid, we have shown how their combination can add considerable extra \onreview{information} on open cosmological questions about the nature of dark energy, masses of neutrinos, and the background curvature of spacetime. \onreview{The overall constraining power across parameters is typically a factor two to five weaker than for the \textit{Euclid} experiment, but this level of precision will still provide vital independent verification of the cosmological model as we move to a paradigm in which systematic uncertainties become comparable or dominate over purely statistical ones. Multiple previous analyses have shown how the next Phase of the SKAO telescopes will be capable of even more, going beyond \textit{Euclid} and other Stage IV experiments.} Furthermore, \onreview{even in the AA4 era,} individual parts of these \twopoint cross-correlations will also be invaluable for other science cases which either seek to understand, or rely on understanding, the formation and evolution of radio galaxies and neutral Hydrogen -- calibrating nuisance parameters using both internal self-calibration and external data from world-leading CMB and optical surveys.

\section*{Author Ordering}
Authors are ordered in two sections. The first in order of contribution, the second alphabetically.

\section*{Acknowledgements}
We thank Santiago Casas for initial discussions and advocating for the inclusion of this Chapter. ZS acknowledges support from the research projects PID2021-123012NB-C43, PID2024-159420NB-C43, the Proyecto de Investigación SAFE25003 from the Consejo Superior de Investigaciones Científicas (CSIC), and the Spanish Research Agency (Agencia Estatal de Investigaci\'on) through the Grant IFT Centro de Excelencia Severo Ochoa No CEX2020-001007-S, funded by MCIN/AEI/10.13039/501100011033. JA ackowledges the support of the grant PGC2022-126078NB-C21 funded by MCIN/AEI/10.13039/ 50110001103 and Diputación General de Aragón-Fondo Social Europeo (DGA-FSE) grant 2023-E21-23R funded by Gobierno de Arag\'on. BB-K acknowledges support from INAF for the project
``Paving the way to radio cosmology in the SKA Observatory era: synergies between SKA pathfinders/precursors
and the new generation of optical/near-infrared cosmological surveys'' (CUP C54I19001050001). CU was supported by the European Union (ERC StG, LSS\textunderscore BeyondAverage, 101075919). MM acknowledges the support of the INFN project “InDark” and the ASI/LiteBIRD grant n.2020-9-HH.0. 

\bibliographystyle{abbrvnat-maxbibnames4}
\bibliography{chapter-newstyle} 

\end{document}